# Structural and Dielectric Characterization on Multiferroic $xNi_{0.9}Zn_{0.1}Fe_2O_4/(1-x)PbZr_{0.52}Ti_{0.48}O_3$ Particulate Composite

Rishikesh Pandey, Braj Raj Meena and Akhilesh Kumar Singh*

School of Materials Science & Technology, Indian Institute of Technology (Banaras Hindu University) Varanasi- 221005, India

## ABSTRACT

We have carried out the powder x-ray diffraction and dielectric studies on multiferroic particulate composite $xNi_{0.9}Zn_{0.1}Fe_2O_4/(1-x)PbZr_{0.52}Ti_{0.48}O_3$ with x=0.15, 0.30, 0.45, 0.60, 0.75 and 0.90 to explore the structural and ferroelectric properties. A conventional double sintering method was used to prepare the $xNi_{0.9}Zn_{0.1}Fe_2O_4/(1-x)PbZr_{0.52}Ti_{0.48}O_3$ composites. The structure of one of the component $Ni_{0.9}Zn_{0.1}Fe_2O_4$ is spinel cubic with space group $Fd\bar{3}m$, while the other component $PbZr_{0.52}Ti_{0.48}O_3$ is selected around the morphotropic phase boundary region in which the tetragonal and monoclinic phases with space group P4*mm* and C*m* coexist respectively. We have carried out Rietveld refinement of the structure to check the formation of ideal composites with separate ferroelectric and ferrite phases. Even though the structural characterization does not reveal the formation of any new phase due to reaction between the two components of the composite during sintering, the tetragonality of the $PbZr_{0.52}Ti_{0.48}O_3$ continuously decreases with increasing the ferrite fraction while the lattice parameter of ferrite phase increases with increasing fraction of the ferroelectric phase. Similarly, the dielectric study reveals clear shift in the ferroelectric to paraelectric phase transition temperature of $PbZr_{0.52}Ti_{0.48}O_3$ during composite formation suggesting that part of $Ni^{2+}$, $Zn^{2+}$/ $Fe^{3+}$ ions are diffusing at the B-site of $PbZr_{0.52}Ti_{0.48}O_3$ replacing $Ti^{4+}$, which in turn decreases its transition temperature. Scanning electron micrograph of sintered pellet surface confirms the presence of two types of particle morphology in the particulate composite, corresponding to ferrite and ferroelectric phases.

* Corresponding author

E-mail address: akhilesh_bhu@yahoo.com, aksingh.mst@itbhu.ac.in

## 1. INTRODUCTION:

In recent years, multifunctional materials such as multiferroics have stimulated great interest of researchers due to the potential applications for devices such as sensors, actuators, transducers and memories [1-2]. In multiferroics, more than one ferroic properties like ferroelectricity (FE), ferromagnetism (FM) etc. are present simultaneously and are coupled. For example, presence of coupled ferroelectric and ferromagnetic responses provides an extra degree of freedom in which data could be written electrically and read magnetically and vice versa for memory applications [3]. Usually presence of ferroelectricity in perovskites requires the empty d-orbitals at B-site cation which soften the orbital overlapping between B-site cations and O anions in the $ABO_3$ perovskite structure [4]. However, ferromagnetism requires unpaired electrons in the d-orbitals of the ions at B-site. That is the reason why these materials are rarely found in nature and difficult to synthesize in laboratory [5]. Single phase multiferroics such as $BiFeO_3$, $BiMnO_3$, etc. exhibit very low magnetoelectric (ME) coefficient and low resistivity limiting their applications [6-7]. Most of them exhibit magnetoelectric coupling much below room temperature and therefore are not suitable for potential device applications [3, 8-9]. In view of this, the composite multiferroics in which two phases are coupled via strain may be a good substitute for single phase multiferroics [2]. The composite multiferroics may exhibit high ME- coefficient which have not been observed in the single phase ME materials. The appearance of magnetoelectric coupling in composites is relatively new concept introduced in 1972 by Suchtelen [10]. Subsequently large number of papers appeared, reporting significant ME-coefficient (~ 100 V/cm-Oe) in ferroelectric/ ferromagnetic composites [11]. In the recent years, large numbers of particulate composites and piezoelectric-magnetostrictive heterostrucures have been explored such as $Ni_{0.9}Co_{0.1}Fe_2O_4/Pb(Zr_xTi_{1-x})O_3$ [12], $BiFe_{0.5}Cr_{0.5}O_3/NiFe_2O_4$ [13], $Ni_{0.93}Co_{0.02}Mn_{0.05}Fe_{1.95}O_4/$ $Pb(Zr_xTi_{1-x})O_3$ [14], $Pb(Zr_{0.52}Ti_{0.48})O_3/NiFe_2O_4$ [15], $BaTiO_3/(Ni_{0.3}Zn_{0.7})Fe_{2.1}O_4$ [16], $Pb(Zr_{0.52}Ti_{0.48})O_3/NiFe_{1.9}Mn_{0.1}O_4$ [17], $BaTiO_3/NiFe_{1.98}O_4$ [18], $CoFe_2O_4/PbZr_{0.52}Ti_{0.48}O_3$ [19] etc. Significant improvement in ME-coefficient (~ 25 mV/cm-Oe) is reported in the composites prepared by $NiFe_2O_4$ (NFO) and $Pb(Zr_xTi_{1-x})O_3$ (PZT) using spark plasma sintering method [20]. Particularly the composite heterostructure of Terfenol-D with PVDF prepared in the form of multilayer thin film which exhibit very high ME-coefficient (> 1 V/cm-Oe) termed as giant magnetoelectric coefficient (GME) [21]. There are several papers in the literature which report the magnetoelectric investigations on particulate composites using various types of ferroelectric phases such as $BaTiO_3$, PZT etc. and magnetic phases such as nickel ferrite, cobalt ferrite, nickel zinc ferrite etc. assuming that the two components form ideal composite without reacting/ modifying each other. However, at high sintering temperature during composite formation it is very unlikely that the magnetic and ferroelectric phases remain intact without diffusion of ions from one component to other. To the best of our knowledge the possible modifications in the ferroelectric and magnetic components during particulate composite formation has not been investigated into detail which may have very crucial impact on the ME response of the composite. In the present work we have looked into this aspect by investigating several compositions of $xNi_{0.9}Zn_{0.1}Fe_2O_4/(1-x)PbZr_{0.52}Ti_{0.48}O_3$ (NZFO/PZT) composite.

Multiferroic composite of NZFO and PZT has been prepared choosing the suitable composition of perovskite PZT in the MPB region and NZFO using conventional double sintering route. PZT is a well-known piezoelectric material and the composition $Pb(Zr_{0.52}Ti_{0.48})O_3$ chosen is in the MPB region in which both tetragonal (P4*mm*) and monoclinic (C*m*) phases coexists showing maximum piezoelectric response [22]. The transition temperature for PZT with x=0.52 is ~ 380 $^{0}C$ [23]. Nickel zinc ferrite ($Ni_{0.9}Zn_{0.1}Fe_2O_4$) shows ferrimagnetic to paramagnetic phase transition at $T_N$ ~ 530 $^{0}C$ [24]. Our study reveals that both the ferrite and ferroelectric phases get modified during composite formation at high sintering temperature which in turn may deteriorate the ME response of the composite. We observed a systematic variation in the lattice parameter of the ferrite and ferroelectric phases with changing their phase fraction in the composite which suggest that the ions from the two components are diffusing into each other thereby modifying the lattice parameters. The transition temperature ($T_C$) of PZT also shows a systematic shift to lower temperatures for various compositions of the composite and the ferroelectric to paraelectric transition becomes more diffused with increasing the ferrite fraction in the composite. To prevent the reaction between two components and to improve the ME response of such particulate composites, the sintering temperature needs to be brought down by using nanocrystalline powders [25] or by reducing the sintering time by adopting the methods like hot pressing [26] or spark plasma sintering [20].

## 2. EXPERIMENTAL DETAILS:

Samples used in the present work were prepared by conventional double sintering solid state route. To prepare $Ni_{0.9}Zn_{0.1}Fe_2O_4$, stoichiometric amounts of AR grade NiO (obtained by thermal decomposition of $NiCO_3.2Ni(OH)_2$ (QUALIGENS) at 550 $^{0}C$), ZnO (QUALIGENS, 99%) and $Fe_2O_3$ (HIMEDIA, 99%) were mixed in agate mortar and then ball milled for 4 h. The powder mixture was calcined at 800 $^{0}C$ for 6 h. Similarly, $PbZr_{0.52}Ti_{0.48}O_3$ (PZT) was prepared by conventional solid state route using AR grade $PbCO_3$ (HIMEDIA, 99.9%), $ZrO_2$ (HIMEDIA, 99%) and $TiO_2$ (HIMEDIA, 99%). Stoichiometric amounts of these ingredients were ball milled in acetone (as mixing media) for 6 h. The mixed powder was dried and then calcined at 800 $^{0}C$ for 6 h. Powder x-ray diffraction (XRD) patterns were recorded using an 18 kW rotating Cu-target based RIGAKU (Japan) powder x-ray diffractometer fitted with a graphite monochromator in the diffracted beam. To prepare different compositions of NZFO/PZT composites, the mixture of calcined powders of PZT and NZFO were first ball-milled in acetone for 6 h and then dried. 2% polyvinyl alcohol (PVA) solution in water, which acts as binder was mixed with the powders. The powder mixture was pressed in the form of pellets of diameter 12 mm and thickness ~ (1.5- 2.0) mm using a stainless- steel die and uniaxial hydraulic press at an optimized load of 65 kN. Before sintering, the green pellets were kept at 500 °C for 10 h to burn out the binder. The pellets were finally sintered at 1150 $^{0}$C for 6 h in PbO atmosphere in sealed Alumina crucible. For dielectric measurements, the flat surface of sintered pellets was gently polished with 0.25 μm diamond paste for about 2 min and then washed with acetone. Isopropyl alcohol was then applied for removing the moisture, if any, on the pellet surfaces. Fired on silver paste was subsequently applied on both the surfaces of the pellets. It was first dried around 120 $^{0}$C in an oven and then cured by firing at 500 $^{0}$C for about 5 min. Dielectric measurements were carried out using a Novocontrol, Alpha-A high performance frequency analyzer. For the high temperature dielectric measurements, the temperature of the sample was controlled by using a Eurotherm programmable temperature controller with an accuracy of ±1 $^{0}$C. The measurements were carried out during heating the sample at a rate of 1 $^{0}$C per min. FULLPROF program (Rodriguez- Carvajal) [27] was used for Rietveld refinement of the crystal structure. Pseudo-Voight function was used to define the peak profiles and sixth-order polynomial was used to fit the background. In the spinel cubic phase of space group $Fd\bar{3}m$ (Space group # 227), $Zn^{2+}$ and $Fe_I^{3+}$ ions occupy the tetrahedral 8(a) sites at (1/8, 1/8, 1/8), $Fe_{II}^{3+}$ and $Ni^{2+}$ ions occupy octahedral sites 16(d) at (1/2, 1/2, 1/2) and $O^{2-}$ ions occupy 32(e) sites at (-x+1/4, -x+1/4, x) as listed in the "International Table for Crystallography, Vol. A (2005)". In the tetragonal phase with P4*mm* space group, the $Pb^{2+}$ ion occupies 1(a) sites at (0,0, z), $Ti^{4+}$/$Zr^{4+}$ and $O_I^{2-}$ occupy 1(b) sites at (1/2, 1/2, z), and $O_{II}^{2-}$ occupy 2(c) sites at (1/2, 0, z). In the monoclinic phase with space group C*m*, $Pb^{2+}$, $Ti^{4+}$/ $Zr^{4+}$ and $O_I^{2-}$ occupy 2(a) sites at (x, 0, z) and $O_{II}^{2-}$ occupy 4(b) sites at (x, y, z). The microstructure of the sintered pellets was studied using scanning electron microscope (SEM) (ZEISS SUPRA 40). Before the microstructural study, sintered pellets were sputter coated with Pd/Au alloy.

## 3. RESULTS & DISCUSSION:

### 3.1 Crystal Structure:

The XRD patterns of calcined powders of NZFO and PZT are shown in Fig.1(a) and (b), respectively. All the peaks shown in Fig.1(a) are indexed with the cubic spinel structure with space group $Fd\bar{3}m$ and no impurity phase is present. Similarly, for PZT all the peaks shown in Fig.1(b) correspond to perovskite structure. The indices shown on the peaks in Fig.1(b) correspond to pseudocubic cell. For this composition of PZT, both the tetragonal (P4*mm*) and monoclinic (C*m*) phases are reported to coexist [22] which is evident from the broad triplet character of the (200) pseudocubic profile. The Rietveld refinement of the structure by us confirms the coexistence of the monoclinic and tetragonal structures. Fig.2 shows the XRD profile for different compositions of NZFO/PZT composites prepared by sintering the mixture of the calcined powders of NZFO and PZT in different proportions. Peaks marked with "p" and "n" denote the reflections corresponding to PZT and NZFO phases, respectively. We can see that as we increase the fraction of NZFO, the intensity of peaks corresponding to PZT decreases while the peak intensity corresponding to NZFO increases. For all the patterns shown in Fig.2, except the peaks corresponding to the PZT and NZFO, there is no additional peak. This suggests that during composite formation any new phase due to possible reactions between the two components is not present. Phase fractions of the PZT and NZFO were calculated using the intensity of the most intense peak of the XRD pattern of the two phases using equation (i) and (ii),

$$\text{Phase fraction of NZFO} = \frac{I_{(311)\mathrm{NZFO}}}{(I_{(311)\mathrm{NZFO}} + I_{(110)\mathrm{PZT}})} \quad \text{(i)}$$

$$\text{Phase fraction of PZT} = \frac{I_{(110)\mathrm{PZT}}}{(I_{(311)\mathrm{NZFO}} + I_{(110)\mathrm{PZT}})} \quad \text{(ii)}$$

As shown in Fig.3(a) and Fig.3(b), we obtained good agreement between this calculated phase fractions (%) with weight fraction of the PZT and NZFO used to prepare various compositions. This again suggests that the particulate composite formation during sintering of the mixture is not leading to any new phase.

We could successfully refine the structure of the various compositions of the composites using Rietveld method. The Rietveld fit for the XRD pattern of NZFO/PZT composite with x=0.60 is shown in Fig.4. Using the cubic space group $Fd\bar{3}m$ for NZFO, and coexisting tetragonal (P4*mm*) and monoclinic (C*m*) space groups for PZT, very good fit between observed and calculated profiles is obtained. The refined structural parameters for pure NZFO, PZT and NZFO/PZT composite with x=0.60 are listed in table I. The isotropic thermal parameters ($B_{iso}$) for the $Pb^{2+}$ ion was found to be very high (>3 $Å^2$). In view of this, we considered anisotropic thermal parameters (β) for these ions in tetragonal and monoclinic structures. For the cubic phase, $B_{iso}$ were used for all the ions. The refined structural parameters for pure PZT are in good agreement with that reported by Ragini et al [22].

Variation of lattice parameter of NZFO and tetragonal phase of PZT obtained after Rietveld refinement of the structures for various compositions of composite is shown in Fig.5(a) and (b) respectively. As can be seen from Fig.5(a) the lattice parameter of NZFO systematically increases with increasing the fraction of PZT in the composite. Similarly the 'a' and 'c' parameters of the tetragonal phase of PZT come closer to each other with increasing the fraction of the NZFO in the composite. The tetragonality (c/a) continuously decreases with increasing the fraction of NZFO and structure becomes pseudocubic for x=0.85. These modifications in the lattice parameters of NZFO and PZT are not possible unless the ions from the two components diffuse to each other. Thus the solid state sintering route of the composite formation is not leading to the ideal composite where the two components should not react with each other [28]. Most important impact of the modification in the lattice parameter will correspond to the ferroelectric phase where reduced tetragonality will decrease the magnitude of the polarization of the unit cell which in turn will reduce the piezoelectric response of the ferroelectric phase. Since the ME effect in particulate composite results from the coupling between polarization and magnetization mediated by strain, the reduced piezoelectric response will decrease the value of ME coefficient also.

### 3.2 Microstructure studies:

The SEM image for NZFO/PZT composite with x=0.60 is shown in Fig.6. Appearance of two types of morphology could be seen clearly in the SEM image confirming the formation of composite. The smaller grains with elongated oval shape are identified as NZFO. The rock shaped bigger grains correspond to PZT. The average grain size for NZFO and PZT is calculated to be ~ 0.09 μm and 1.0 μm, respectively. As can be seen from the micrograph, there is uniform distribution of the two components.

### 3.3 Dielectric Studies:

The room temperature permittivity ($\varepsilon^{/}$) of the NZFO/PZT composite in the frequency range of (1 kHz- $10^6$ kHz) for different composite compositions is shown in Fig.7. It is evident from Fig.7 that with increasing the amount of NZFO, the permittivity decreases. This is due to decreasing the fraction of the ferroelectric phase PZT. Another reason for decrease in the dielectric constant with increasing ferrite content may be attributed to the Verwey type electron exchange polarization [29].

Fig.8 shows the temperature variation of permittivity for NZFO/PZT composite with x=0.60 at various frequencies. With increasing frequency, the peak in the temperature variation of the permittivity shifts to the higher temperature side along with the significant frequency dispersion. This suggests the relaxor nature of the ferroelectric phase transition in PZT after composite formation. However, the pure PZT behaves like a normal ferroelectric and does not exhibit relaxor features [30]. The appearance of relaxor features must be linked with the modification of the PZT phase by diffusion of ions from NZFO during composite formation. It is well known that the relaxor nature of ferroelectric phase transition is associated with the local ionic size/charge disorder in perovskite solid solutions [31]. Relaxor nature of ferroelectric phase transition in the NZFO/PZT composite must be resulting from the introduction of more disorder at the B-site cations in the $ABO_3$ perovskite structure due to diffusion of $Ni^{2+}/Zn^{2+}/Fe^{3+}$ ions from the NZFO to the $Zr^{4+}/Ti^{4+}$ site. This may lead to the Maxwell-Wagner type interfacial polarization [32] as discussed in Koop's phenomenological theory [33].

Fig.9 shows the temperature variation of permittivity at a frequency of 10 kHz for the NZFO/PZT composite with x=0, 0.15, 0.30, 0.45 and 0.60. For pure PZT with x=0 the permittivity shows a peak at 380 $^0$C which is in well agreement with the value of $T_C$ reported in literature [23] for $Pb(Zr_{0.52}Ti_{0.48})O_3$. With increasing the value of 'x' i.e. the NZFO fraction the permittivity peak systematically shifts to the lower temperature side and the nature of phase transition becomes very diffused. This suggests that the relaxor feature of the phase transition become more pronounced with increasing NZFO concentration in the composite. As discussed earlier, the appearance of relaxor features is linked with the diffusion of $Ni^{2+}/Zn^{2+}/Fe^{3+}$ ions from NZFO to the $Zr^{4+}/Ti^{4+}$ site of the PZT. It is expected that at higher 'x' more $Ni^{2+}/Zn^{2+}/Fe^{3+}$ ions are available to diffuse to the $Zr^{4+}/Ti^{4+}$ site of the PZT which in turn strengthen the relaxor nature of the ferroelectric phase transition. To elaborate the strengthening of the relaxor character with increasing NZFO concentration in the composite, we have plotted the full width at half maximum (FWHM) of the permittivity peaks at 10 kHz for various compositions in Fig.10. It is evident from Fig.10 that the FWHM increases continuously with increasing NZFO concentration, indicating enhanced diffuseness of the ferroelectric phase transition. Fig.11 shows the composition dependence of $T_C$ determined at 1 kHz frequency for NZFO/PZT composite. As compared to pure PZT with $T_C$=380 $^0$C, the $T_C$ decreases drastically for x=0.15 (~ 350 $^0$C) and then decreases linearly with increasing NZFO concentration. The significant change in $T_C$ of PZT after composite formation clearly suggests that the ferrite and ferroelectric phases react with each other. This is in contrast to the proposition of the Chougule et al. [34] that the two phases do not react.

Chougule et al. [34] have reported that the transition temperature for the composition with x=0 (for pure $PbZr_{0.52}Ti_{0.48}O_3$) is 420 $^0$C, which is incorrect, and not consistent with the PZT phase diagram [23]. Our study on the composition with x=0 shows that the $T_C$ ~ 380 $^0$C which is consistent with the reported phase diagram of PZT for pure PZT composition in the MPB region [23,35]. Similarly, for the other compositions with x=0.15, 0.30 and 0.45, the Chougule et al. have reported that $T_C$ is 380 $^0$C, 375 $^0$C and 370 $^0$C. In contrast, our studies clearly reveal that the $T_C$ for x=0 is 380 $^0$C, decreases drastically to ~ 350 $^0$C for x= 0.15, and then shows linear decrease for x=0.30, 0.45 and 0.60 as 348 $^0$C, 345 $^0$C and 342 $^0$C, respectively. Also, the reported value of $\varepsilon^/$ (= 6160) by Chougule et al. [34] for pure NZFO is extremely high and comparable to ferroelectric (PZT) rich composition with x=0.15 ($\varepsilon^/$= 6446). Such high value of $\varepsilon^/$ for NZFO is not expected. The reported value of $\varepsilon^/$ in literatures is as high as ~ 850 (at 1 kHz) at the sintering temperature of 800 $^0$C which falls down to ~ 250 with increasing the sintering temperature above 1000 $^0$C [36-37]. The value of $\varepsilon^/$ (~ 200) measured at room temperature and 1 kHz by us (see Fig.7), for pure NZFO sintered at 1150 $^0$C is nearly comparable to that reported by Chen et al [36].

## 4. CONCLUSIONS:

The dielectric and structural characterization of NZFO/PZT composites by us reveals that at the XRD level no secondary phases appear after composites formation. However, the lattice parameters of both the ferrite and ferroelectric phases are modified suggesting that the ions from the two phases diffuse to each other during sintering. The tetragonality of ferroelectric phase continuously decreases with increasing the ferrite fraction in the composite. To get the maximum piezoelectric response of the ferroelectric phase and consequently better ME-coefficient in composite, a composition with higher tetragonality may be chosen, the structure of which latter may get modified to MPB phase. The transition temperature of the ferroelectric phase (PZT) also decreases significantly after composites formation and the nature of the phase transition becomes relaxor like. The SEM studies reveal the two types of particle morphology in the particulate composite, corresponding to ferrite and ferroelectric phases.

**ACKNOWLEDGEMENTS:**

RP acknowledges University Grant Commission (UGC), India for financial support. Authors are thankful to Professor Dhananjai Pandey, School of Materials Science and Technology, IIT (BHU) Varanasi, India for extending laboratory facilities and support.

**FIGURE CAPTIONS:**

**Fig.1**Powder XRD pattern of NZFO and PZT ceramics calcined at 800 $^0$C for 6 h.

**Fig.2**Powder XRD profile of xNZFO/(1-x)PZT particulate composites for the composition with x= 0, 0.15, 0.30, 0.45, 0.60, 0.75, 0.90 and 1.0 sintered at 1150 $^0$C for 6 h.

**Fig.3**Variation of phase fractions (%) for PZT ($P_{PZT}$) and NZFO ($P_{NZFO}$) in xNZFO/(1-x) PZT particulate composite.

**Fig.4**Observed (dots), calculated (continuous line), and difference (bottom line) profiles obtained after the Rietveld refinement of the structure of 0.60NZFO/0.40PZT particulate composite sintered at 1150 $^0$C using coexisting [Cubic ($Fd\bar{3}m$)+ Tetragonal (P4*mm*)+ Monoclinic (C*m*)] phases. The vertical tick marks above the difference plot show the positions of the Bragg peaks.

**Fig.5**Variation of the (a) lattice parameter of NZFO in the term of doped PZT (%) in NZFO, and (b) lattice parameters and tetragonality of PZT (tetragonal phase) in the term of doped NZFO (%) in PZT, in xNZFO/(1-x)PZT particulate composite.

**Fig.6**SEM image of the sintered pellet surface of xNZFO/(1-x)PZT particulate composite for the composition with x=0.60 sintered at 1150 $^0$C for 6 h.

**Fig.7**Room temperature permittivity ($\varepsilon^/$) of xNZFO/(1-x)PZT particulate composite for the compositions with x=0, 0.15, 0.30, 0.45, 0.60, 0.75, 0.90 and 1.0 measured in the frequency range 1 kHz- $10^6$ kHz.

**Fig.8**Frequency dependent permittivity ($\varepsilon^/$) of xNZFO/(1-x)PZT particulate composite for the composition with x=0.60 measured in the frequency range 5 kHz -$10^2$ kHz .

**Fig.9**Temperature variation of permittivity ($\varepsilon^/$) of xNZFO/(1-x)PZT particulate composite at 10 kHz for the compositions with x= 0, 0.15, 0.30, 0.45 and 0.60.

**Fig.10** Full width at half maximum (FWHM) variation of permittivity ($\varepsilon^{/}$) peaks at 10 kHz for the compositions with x= 0, 0.15, 0.30, 0.45 and 0.60 of xNZFO/(1-x)PZT particulate composites.

**Fig.11** Variation of the transition temperature ($T_C$) for the different compositions of xNZFO/(1-x)PZT particulate composite with x= 0, 0.15, 0.30, 0.45 and 0.60 at 1 kHz.

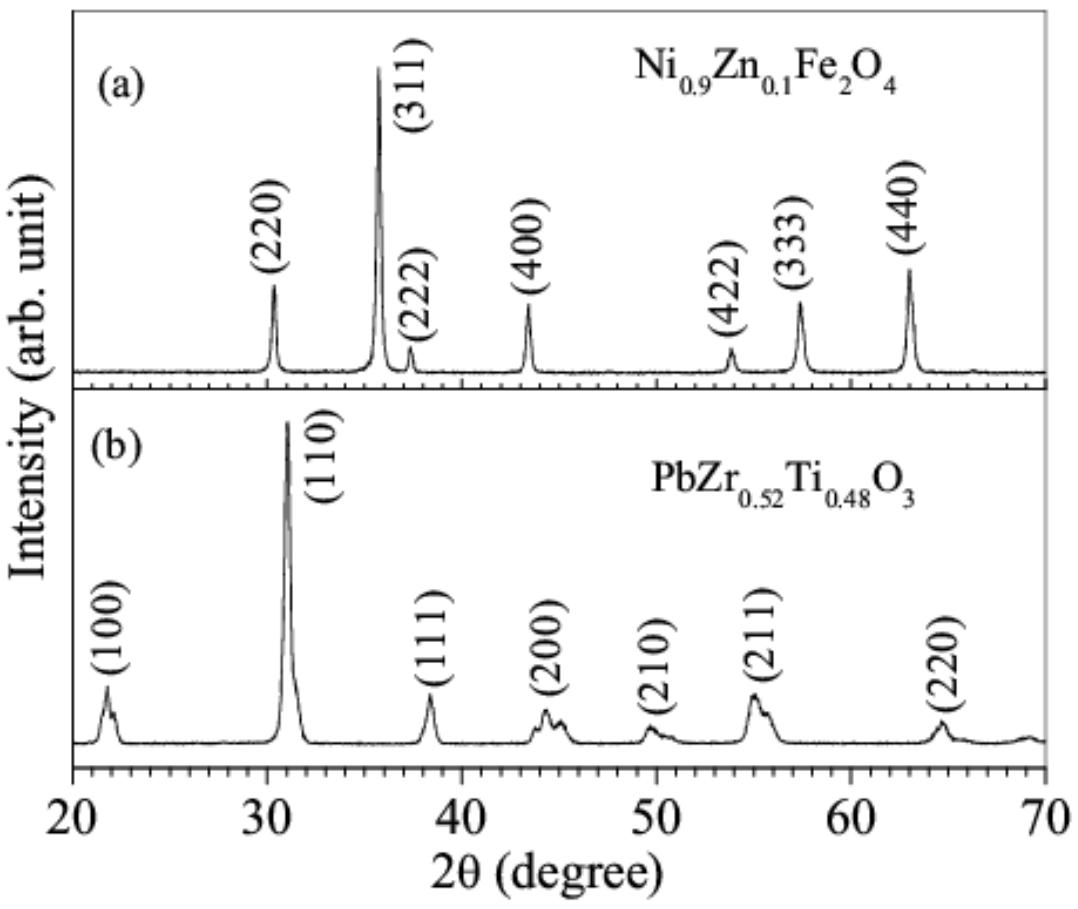

Fig.1

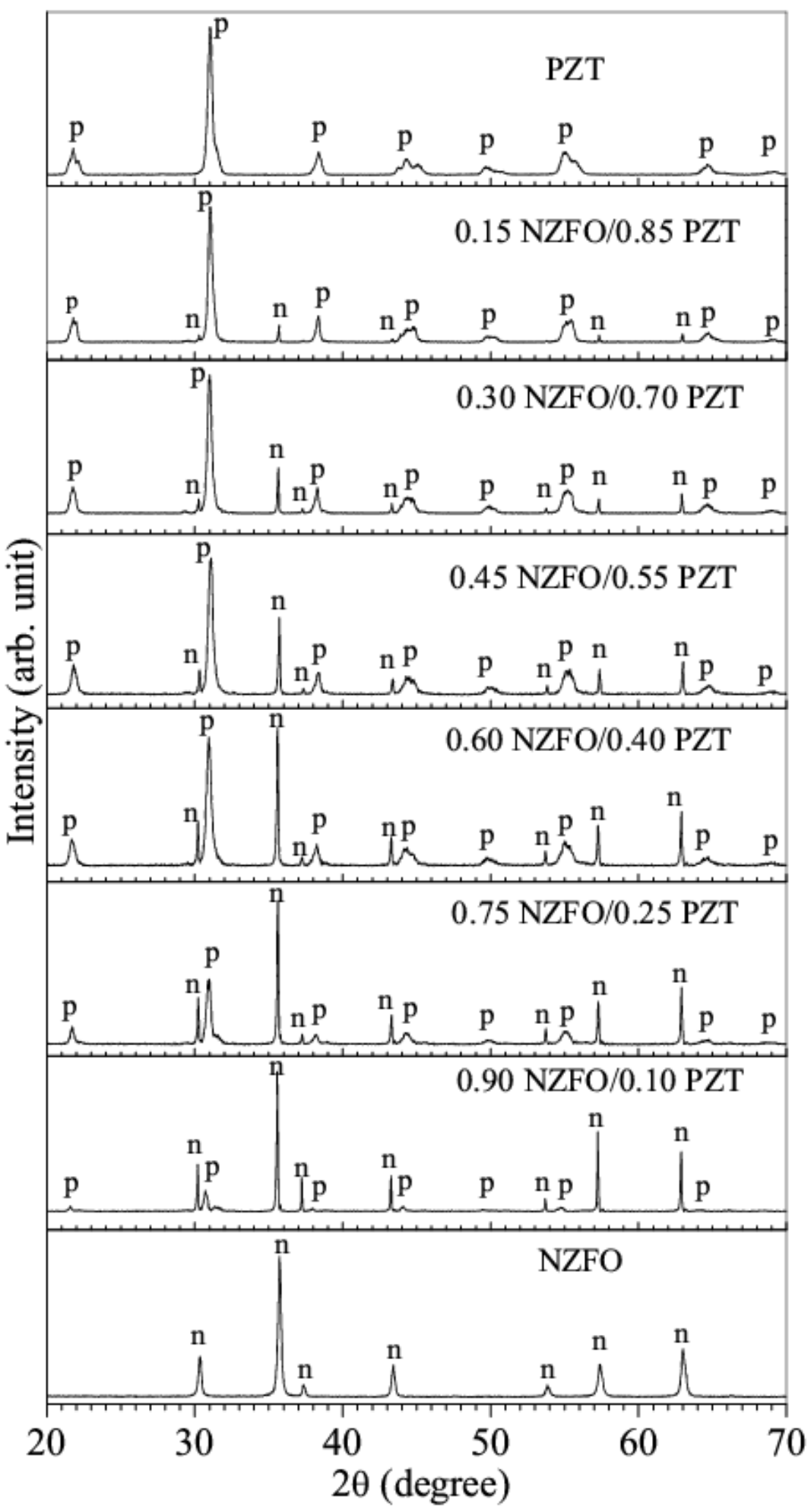

Fig.2

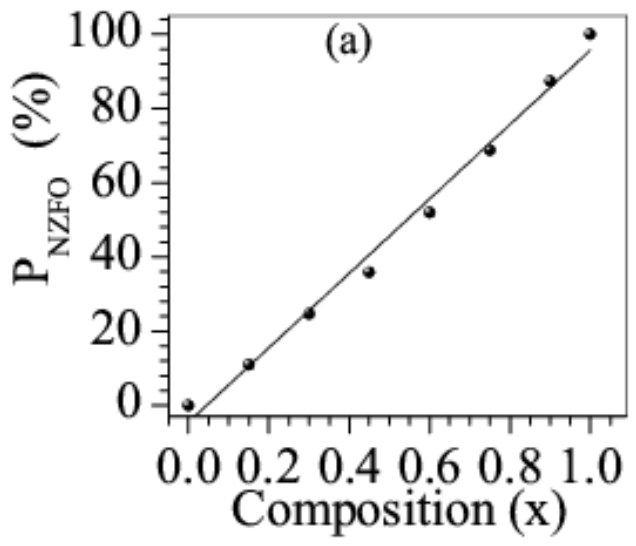

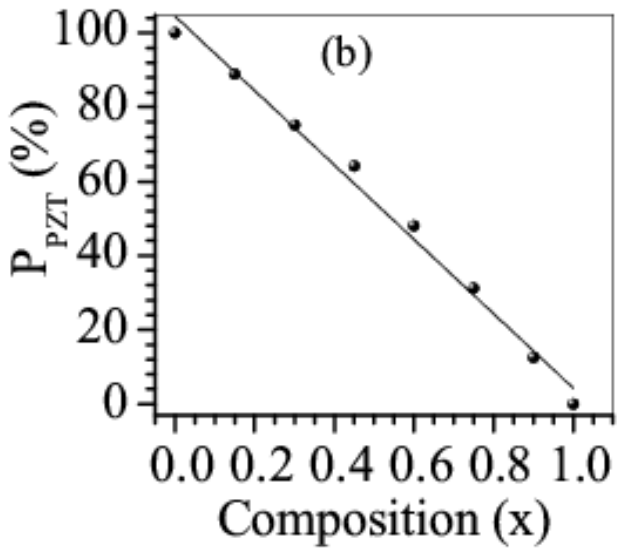

Fig.3

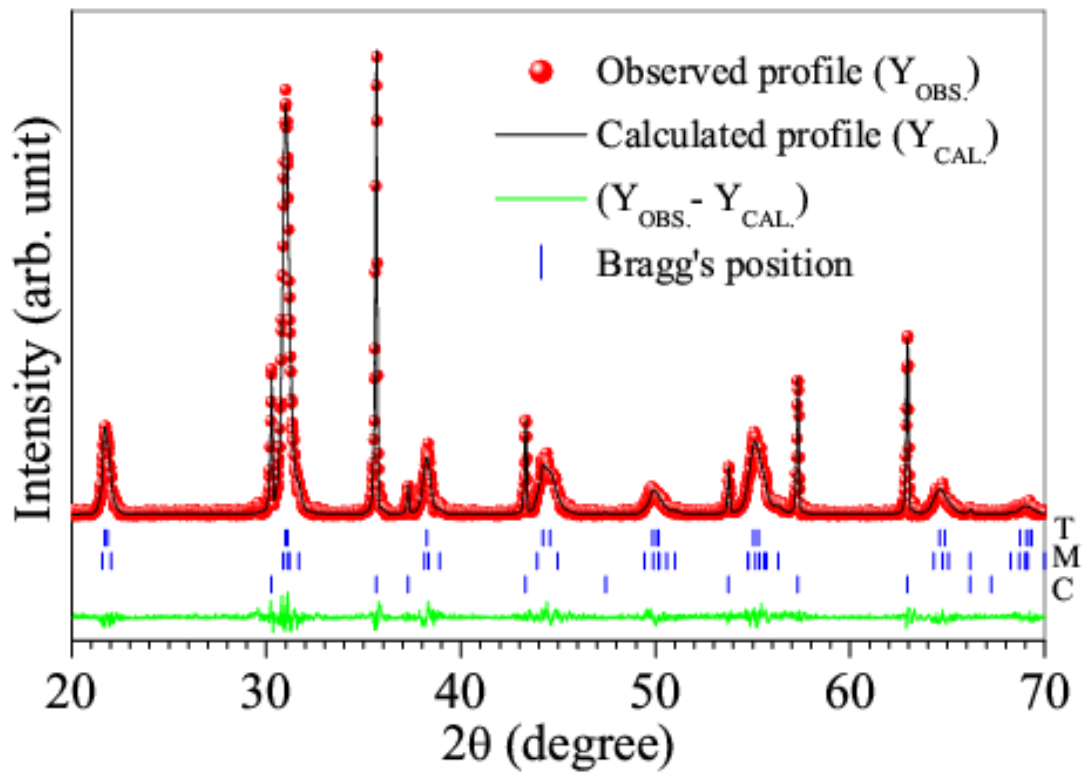
Fig.4

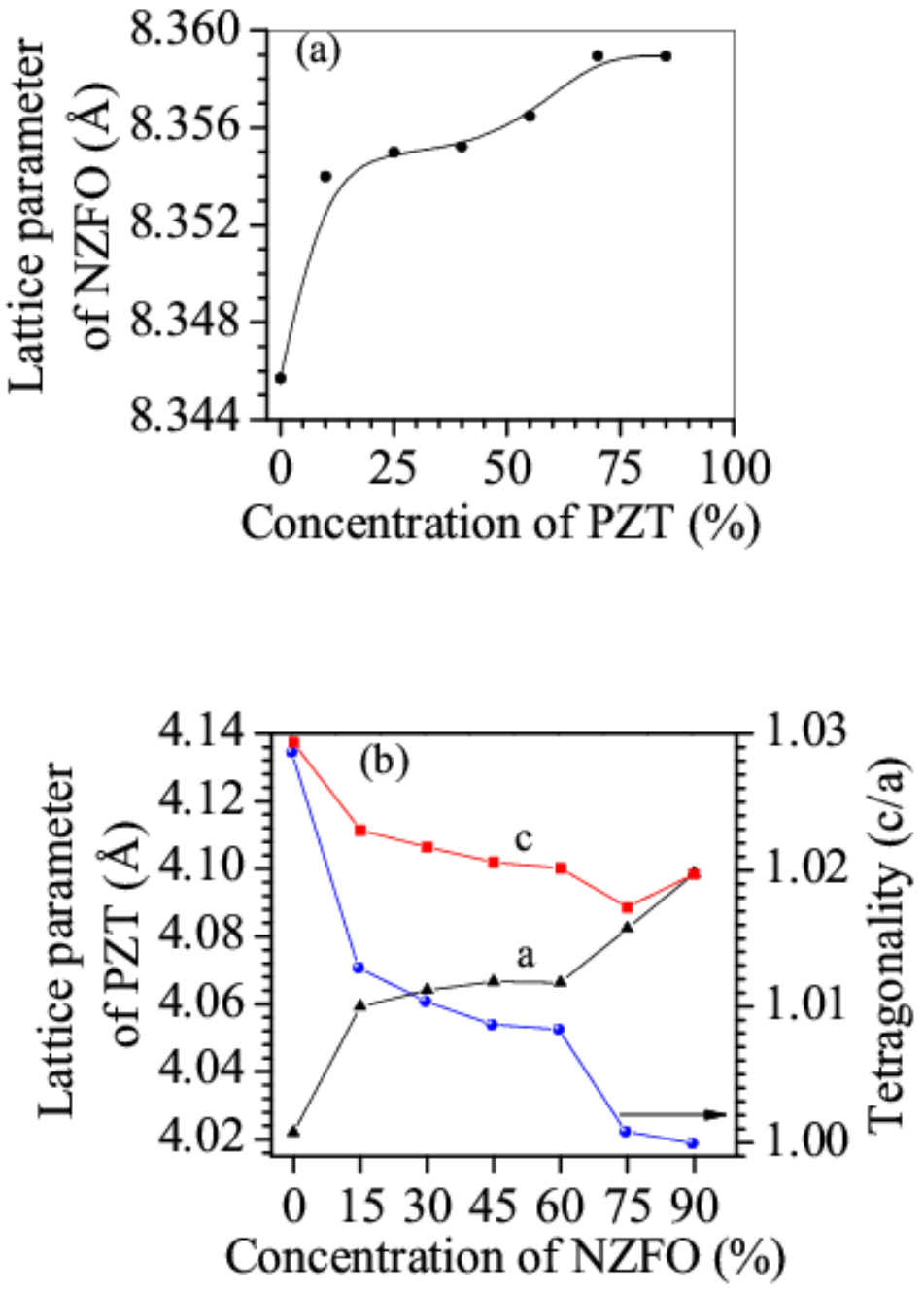

Fig.5

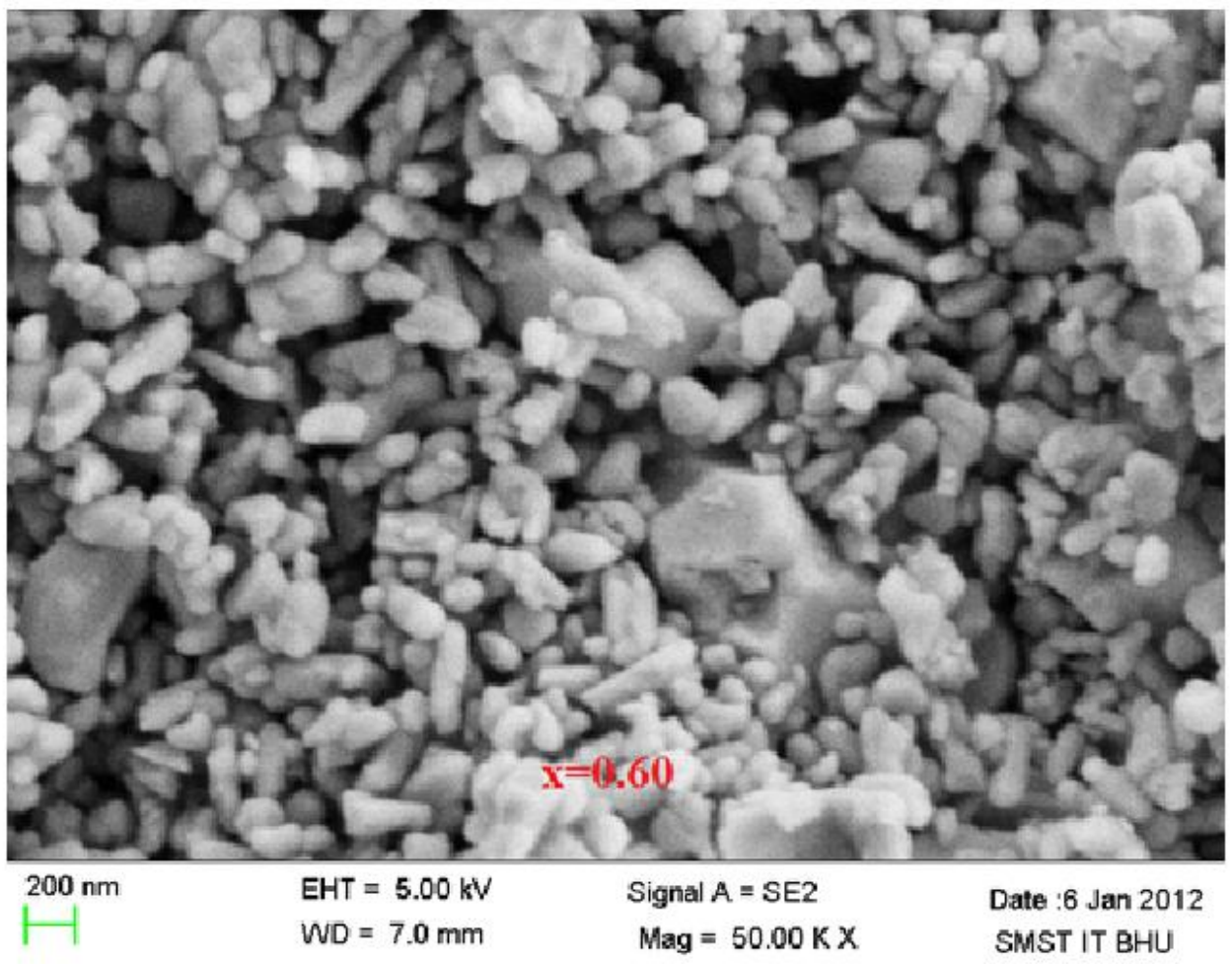
x=0.60
200 nm
EHT = 5.00 kV
WD = 7.0 mm
Signal A = SE2
Mag = 50.00 K X
Date :6 Jan 2012
SMST IT BHU

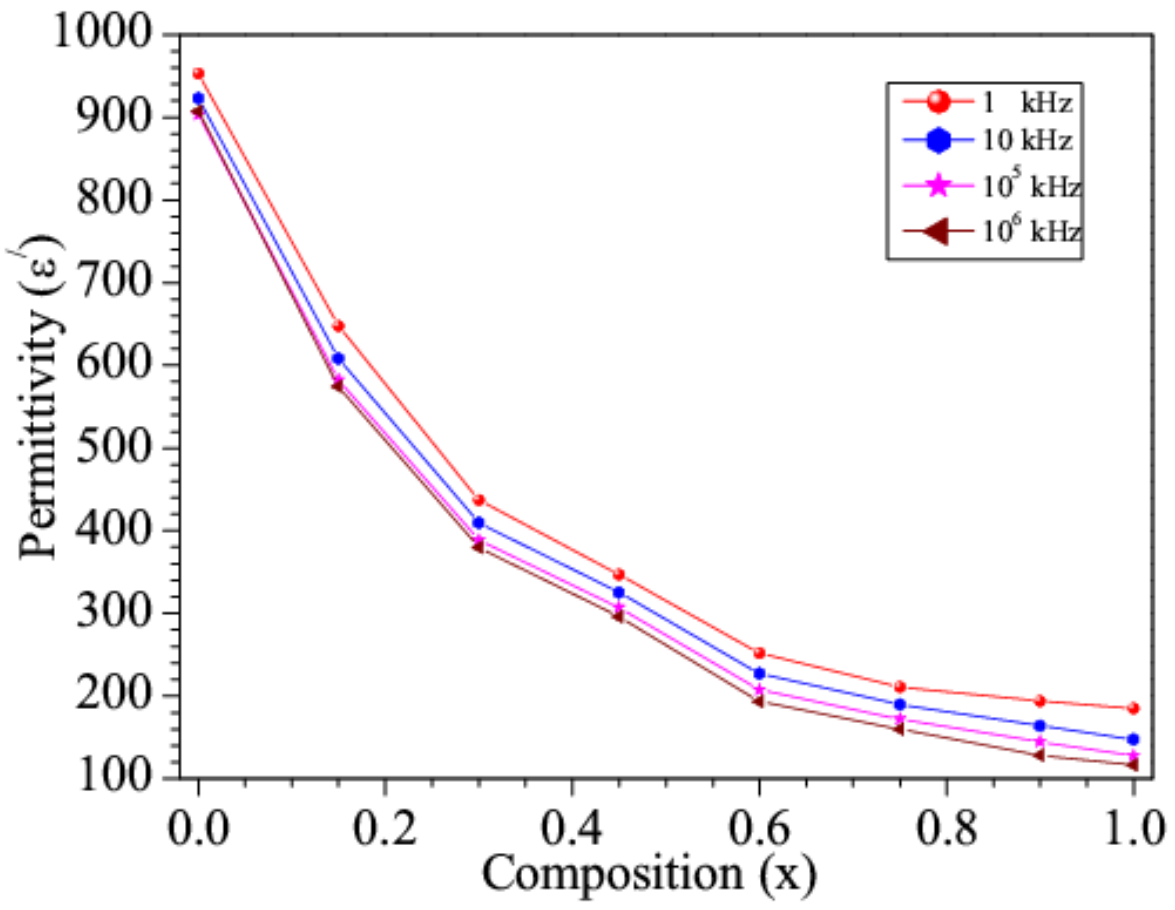

Fig.7

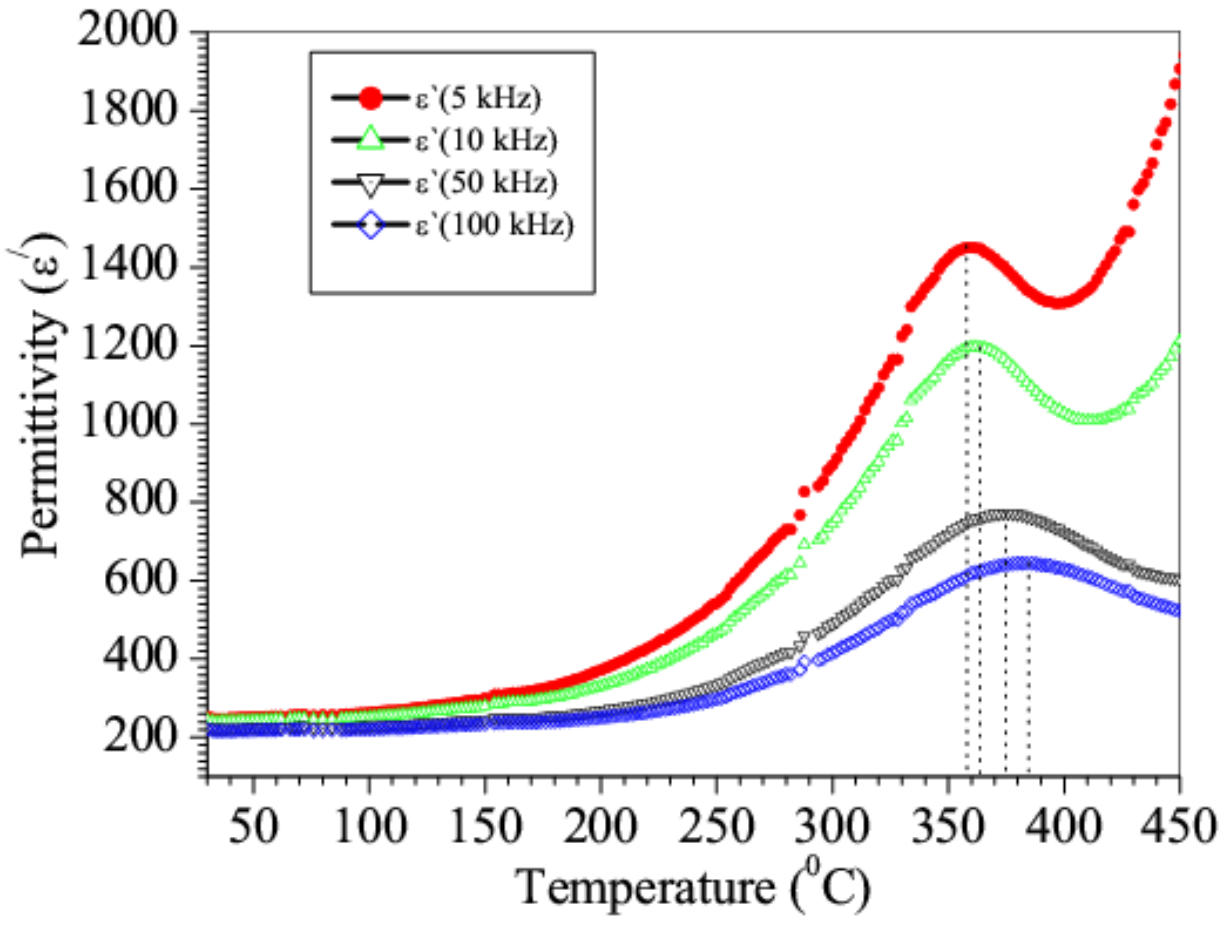

Fig.8

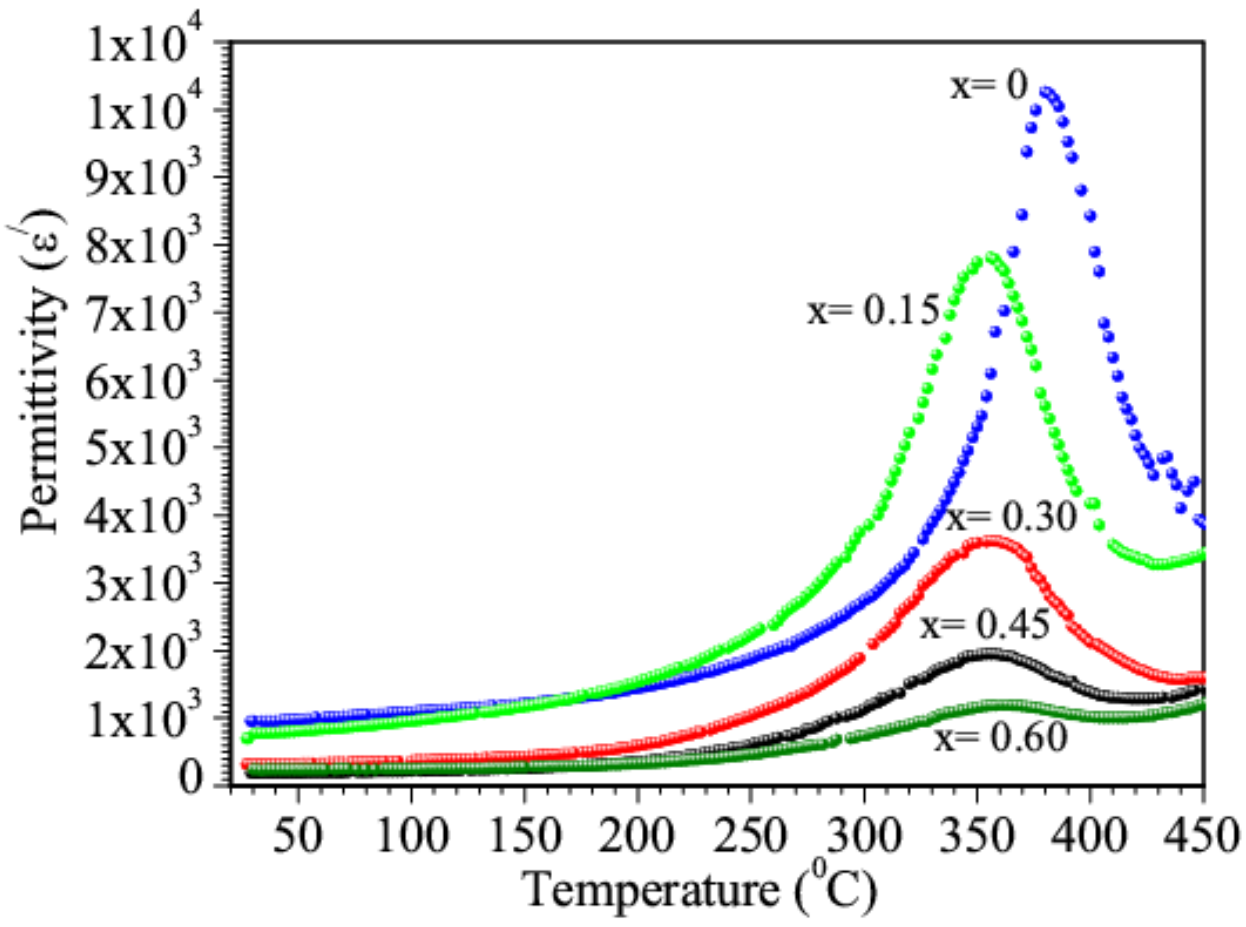

Fig.9

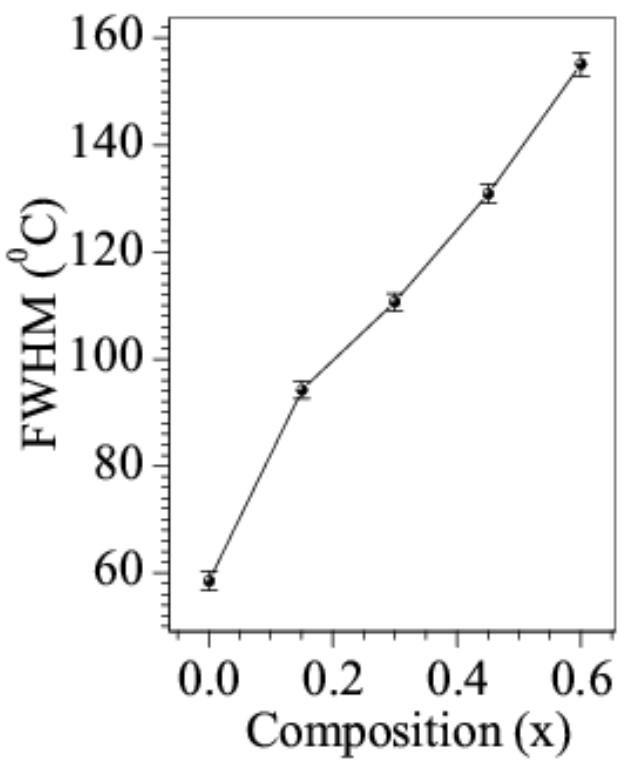

Fig.10

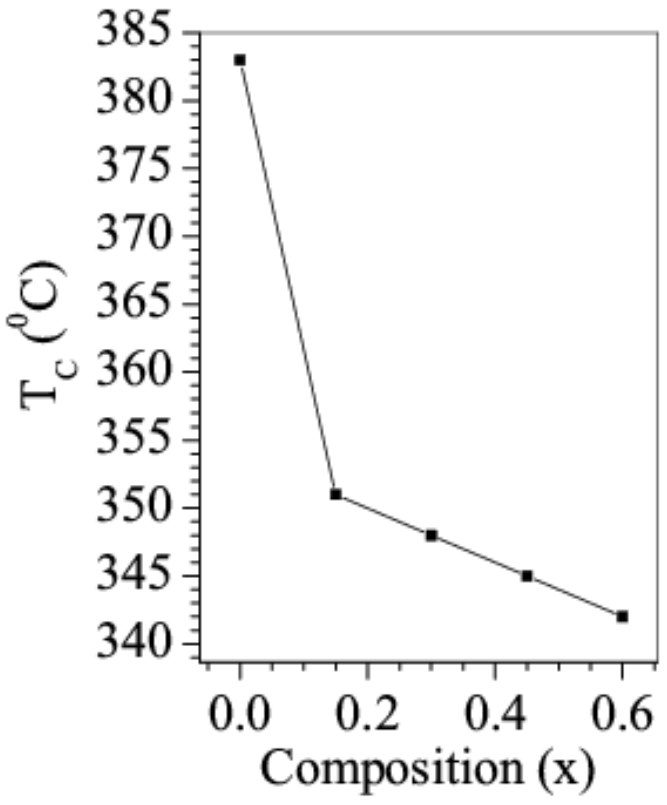

Fig.11

TABLE. 1

| $Ni_{0.9}Zn_{0.1}Fe_2O_4$ ($Fd\bar{3}m$) | $PbZr_{0.52}Ti_{0.48}O_3$ ($P4mm$) | & ($Cm$) |
|---|---|---|
| **Pure** | | |
| $Zn/Fe_I$: x=y=z=0.125,$B_{iso}$=0.59(7) $Å^2$ | Pb:x=y=z=0,$\beta_{11}$=$\beta_{22}$=0.0310(3)$Å^2$,$\beta_{33}$=0.027(3) $Å^2$ | Pb:x=y=z=0,$\beta_{11}$=$\beta_{22}$=0.0310(3)$Å^2$,$\beta_{33}$=0.027(3)$Å^2$ |
| $Fe_{II}$/Ni: x=y=z=0.50, $B_{iso}$= 0.54(6) $Å^2$ | Zr/Ti:x=y=0.50,z= 0.447(2),$B_{iso}$= 0.005(2)$Å^2$ | Zr/Ti:x=0.578(3),y=0,z=0.473(3),$B_{iso}$= 0.0147(3)$Å^2$ |
| O: x=y=z=0.2546(3),$B_{iso}$=1.0(1) $Å^2$ | $O_I$:x=y=0,z=-0.109(6),$B_{iso}$=0.029(1)$Å^2$ | $O_I$:x=0.50(1),y=0,z=-0.10(1),$B_{iso}$=0.01(1)$Å^2$ |
| a=b=c=8.3462(2)Å, $\chi^2$=2.12 | $O_{II}$:x=0.5,y=0.0,z=0.389(3),$B_{iso}$=0.029(1)$Å^2$ | $O_{II}$:x=0.36(1),y=0.219(8),z=0.404(8),$B_{iso}$=0.04(1)$Å^2$ |
| | a=b=4.0429(2)Å,c=4.1318(3)Å, $\chi^2$=1.32 | a=5.752(1)Å,b=5.743(2)Å,c=4.091(4)Å,β=90.48(1)° |
| **Composite** | | |
| $Zn/Fe_I$: x=y=z=0.125,$B_{iso}$=2.2 (2)$Å^2$ | Pb:x=y=z=0, $\beta_{11}$=$\beta_{22}$=0.1204(2)$Å^2$,$\beta_{33}$=0$Å^2$ | Pb:x=y=z=0,$\beta_{11}$=0,$\beta_{22}$=0.0317(2)$Å^2$,$\beta_{33}$=0.045(1)$Å^2$ |
| $Fe_{II}$/Ni: x=y=z=0.50,$B_{iso}$=2.6(2)$Å^2$ | Zr/Ti:x=y=0.50,z=0.5039(2),$B_{iso}$=1.23(1)$Å^2$ | Zr/Ti:x=0.4515(1),y=0,z=0.5281(7),$B_{iso}$=0.3413(1)$Å^2$ |
| O: x=y=z=0.2563(6),$B_{iso}$=3.0(3)$Å^2$ | $O_I$:x=y=0,z=0.1144(7),$B_{iso}$=0.02$Å^2$ | $O_I$:x=0.5924(3),y=0,z=0.0523(7),$B_{iso}$=0.02$Å^2$ |
| a=b=c=8.3552(2)Å | $O_{II}$:x=0.5,y=0,z=0.3933(1), $B_{iso}$=0.03$Å^2$ | $O_{II}$:x=0.312(5),y=.2137(1),z=0.3649(2),$B_{iso}$=0.03$Å^2$ |
| $\chi^2$=4.67 | a= b=4.0663(4)Å, c=4.1001(3)Å, | a=5.657(3)Å,b=5.744(4)Å,c=4.112(3)Å,β=90.92(3)° |